\documentclass{ametsocV6.1}

\usepackage{amsmath}
\usepackage{amssymb}
\usepackage{caption}
\usepackage{subcaption}
\usepackage{svg}
\usepackage{multirow}
\usepackage{floatrow}
\usepackage[normalem]{ulem}
\usepackage{tikz}
\usetikzlibrary{positioning}

\newcommand{\rev}[1]{\textcolor{black}{#1}}

\title{Wave and turbulence separation using dynamic mode decomposition}

\authors{Julio Ch\'avez-Dorado,\aff{a}\correspondingauthor{Julio Ch\'avez-Dorado, jechavez@uw.edu} 
Isabel Scherl,\aff{b}  
and Michelle DiBenedetto,\aff{a} 
}

\affiliation{\aff{a}{Department of Mechanical Engineering, University of Washington, Seattle, WA 98195, USA}\\
\aff{b}{Department of Mechanical \& Civil Engineering, California Institute of Technology, Pasadena, CA 91125, USA}
}

\abstract{Separating the effects of waves and turbulence in oceanographic time series is an ongoing challenge because surface wave motion and turbulence fluctuations can occur at overlapping frequencies. 
Therefore, simple bandpass filters cannot effectively separate their dynamics. While more advanced decomposition techniques have been developed, they often entail restrictive assumptions about the wave and turbulence interactions, require synchronized measurements, and/or only decompose the signal spectrally without a time-series reconstruction. 
We present our new wave-turbulence decomposition technique which uses dynamic mode decomposition (DMD). 
The technique is signal-agnostic so it can be applied to any time series, \rev{and our only assumptions are that the waves and turbulence can be separated and that the waves are the most coherent features in the signal}. 
Our approach requires minimal tuning, where the main user input is the wave frequency range of interest. 
To demonstrate the method, we apply it to synthetic, field, and laboratory data, and compare the results to other mode-based decomposition methods. 
A sensitivity analysis on the synthetic data shows that the most sensitive parameter to the accuracy is the rank truncation in the DMD, \rev{and that the decomposition performs the best when the wave energy in the signal is of equal or greater magnitude than that of the turbulence}. 
Given the accuracy of our decomposition, we are able to analyze the velocity autocorrelation of the separated turbulence time series with minimal wave contamination. 
Overall, our decomposition method outperforms the other decomposition methods and provides for robust separation of the waves and turbulence, demonstrating wide applicability to ocean signal processing.} 

\begin{document}

\maketitle

%
%
%
\statement
    When measuring physical, chemical, and biological quantities in the ocean, the measurements are often influenced by both waves and turbulence. Isolating the individual effects of waves and turbulence on those variables is important in a wide range of analyses, such as estimating how momentum, heat, and nutrients are mixed throughout the water column. In this work, we propose a new method to separate the  wave and turbulence components in ocean data time-series. When tested on laboratory, field, and synthetic data, our method was able to separate the wave and turbulence components of a signal more effectively than previously proposed algorithms.
	
%
%

%
\section{Introduction}

As  observational advances increase the amount of ocean data available \citep{Smith_2019, Rosa_2021}, effectively interpreting and using this information, both in post-processing and real-time analysis, is imperative for gaining insights into the dynamics of the ocean. One ongoing challenge is the interpretation of fluctuating data that is influenced by both turbulence and surface gravity waves.
This is common in data obtained from the ocean surface and from the coastal ocean where both turbulence and surface waves are expected to be strong.
Separating the turbulence and wave fluctuations in a signal is important for a variety of analyses: e.g., isolating the turbulence fluctuations in any signal is necessary for characterizing the turbulence Reynolds stress or for estimating a scalar eddy covariance flux.
And isolating the wave motion is necessary e.g., for quantifying wave energy or for wave energy control strategies \citep{Li_2012, Perez_2020}.
We refer to this separation process as \emph{wave-turbulence decomposition}. 

The difficulties associated with wave-turbulence decomposition lie in the overlapping frequency domains where waves and turbulence exist---and that both can manifest as broadband signals with correlation in space and time. 
Even asserting that these signals can be effectively decomposed is an assumption, given the nonlinear interactions that can occur between waves and turbulence \citep{Jiang_1990, Magnaudet_1995, Guo_2013}.
Over the years, numerous methods have been developed to tackle this problem \citep{Benilov_1974, Jiang_1990, Thais_1995, Trowbridge_1998, Williams_2003, Gerbi_2008, Huang_2008, Young_2018, Bian_2018}. 
However, the various methods are usually adapted to the specific data at hand \citep{Gerbi_2008, Feddersen_2007, Jiang_1990}, involve restrictive assumptions (e.g., \citet{Benilov_1974, Bricker_2007}), and/or require multiple and often complex synchronized measurements \citep{Trowbridge_1998, Doron_2001, Feddersen_2007}.
As a result, we still lack an effective, universal technique for decomposing a \rev{general} time series. 

Our goal is to develop a decomposition method that can be applied to a time series of any type of data. However, for simplicity we will consider a velocity time series as it is the most commonly decomposed signal in this context. Consider the velocity decomposition:
\begin{equation}
    u(x_i,t) = \langle u(x_i,t) \rangle + \tilde{u}(x_i,t) + u'(x_i,t),\label{eq:decomp}
    \end{equation} 
    where $u$ is the instantaneous horizontal velocity, $\langle u \rangle$ is the time average, $\tilde{u}$ is the wave-orbital velocity and $u'$ is the turbulence velocity fluctuation. 
In two dimensions, the result of Reynolds-averaging the Navier-Stokes equations with this decomposition is the Reynolds shear stress:
\begin{equation}
\label{eqn: ReynoldsStress}
   -\frac{\tau}{\rho} = \langle \tilde{u}\tilde{w} \rangle + \langle \tilde{u}w' \rangle + \langle u' \tilde{w} \rangle + \langle u'w' \rangle.
\end{equation}
The first term on the right hand side is identically zero for irrotational waves described by linear wave theory but can be non-zero if the coordinate system is rotated, a common source of corrupted Reynolds stress measurements \citep{Trowbridge_1998}. The second and third terms are zero if there are no interactions between the waves and turbulence. 
The fourth term is the turbulence Reynolds shear stress, which is the only non-zero term as long as the aforementioned assumptions regarding linear wave theory and wave-turbulence interaction hold.

Spectral filtering techniques \citep{Benilov_1974, Bricker_2007, Gerbi_2008, Young_2018} are able to estimate bulk turbulence quantities like the Reynolds stress with some accuracy, but they do not allow for a time-series reconstruction of the decomposed velocity signals because they are applied to the velocity power spectrum.  
Methods based on nonlinear stream functions \citep{Dean_1965, Jiang_1990, Thais_1995,Thais_1996} allow for a time-resolved separation with fewer assumptions but only work for specialized scenarios. 
More recent approaches involve dimensionality-reduction techniques \citep{Huang_1998, Huang_2009, Bian_2018} that make use of mode decomposition to capture the coherent, oscillating nature of waves. However, it is common for these methods to remove turbulence energy in excess from the wave frequency range. In addition, these methods often require tuning parameters that lack a clear physical intuition.
Nevertheless, dimensionality reduction techniques such as proper orthogonal decomposition (POD) \citep{Lumley_1967}, spectral proper orthogonal decomposition (SPOD) \citep{Towne_2018, Schmidt_2020}, and dynamic mode decomposition (DMD) \citep{Rowley_2009, Schmid_2010} have become increasingly popular in fluid dynamics research due to their adaptability and effectiveness in extracting physically interpretable information from complex data. 
For example, POD has been used in pattern recognition in chaotic flows \citep{Albidah_2021}, flow control \citep{Gordeyev_2013}, and unsteady, high-Reynolds-number wake flows \citep{Durgesh_2010}, and SPOD has been employed in diverse flows such as boundary layers \citep{Tutkun_2017}, mixing layers \citep{Braud_2004}, wakes \citep{Araya_2017}, and jets \citep{Heidt_2023}, among others.
Similarly, DMD is widely used in the study of wakes \citep{Schmid_2010}, jets \citep{Schmid_2010, Seena_2011, Schmid_2011a, Semeraro_2012}, instabilities \citep{Duke_2012, Grilli_2012}, and
 oscillations \citep{Seena_2011, Massa_2012, Albidah_2021}.  
 However, these methods have yet to be applied to wave-turbulence decomposition, a problem for which they may be well-suited.
DMD shows promising filtering characteristics for an accurate wave-turbulence decomposition algorithm for single-point velocity measurements because it is designed to capture the time dynamics of coherent structures present in a signal.

Our approach is to develop a decomposition method that is as general as possible. 
We seek a signal-agnostic framework capable of handling one-dimensional time-series data for a broad spectrum of flow parameters. 
Our methodology does not rely on severe assumptions about wave nor turbulence dynamics, making it adaptable to a variety of ocean and environmental datasets.
Thus, we will explore the effectiveness of DMD in a variety of datasets: synthetic data generated with no wave-turbulence interactions, field data collected in a swell-dominated bottom boundary layer, and laboratory data collected in a surface boundary layer under wind-generated waves.
The remainder of the article has the following structure: section \ref{sec:Background} gives relevant background on other wave-turbulence decomposition methods and lays out the mathematical basis for the mode decomposition technique used in our proposed method. Section \ref{sec:Methods} outlines the procedure for our proposed decomposition method. In section \ref{sec:DatasetDescription}, we describe the datasets used for validation, and we present the results of our decomposition in section \ref{sec:ResultsAndDiscussion}, including both a sensitivity analysis and discussion. Finally, we summarize the study in section \ref{sec:Conclusion} and make suggestions for future work.

\section{Background}
\label{sec:Background}

\subsection{Wave-turbulence decomposition}

Wave-turbulence decomposition is a longstanding challenge that has been approached in a variety of ways. 
Simpler methods like moving average and band-pass filters are often used \citep{Foster_1997, Smyth_2002, Williams_2003, Zhu_2016} due to their ease of use and time-resolved output; however, these filters do not distinguish the waves from the turbulence in the overlapping frequency range, resulting in the elimination of turbulent energy.

Many spectral decomposition techniques have been developed that can estimate the Reynolds stress without allowing for a time-series reconstruction; these methods often rely on the correlation between two separate measurements. 
For example, \citet{Benilov_1974} proposed a linear filtration technique that leverages coherence between the pressure and velocity spectra.
Similarly, \citet{Trowbridge_1998} and  \citet{Shaw_2001} proposed a method that leverages the coherence between two synchronized velocity measurements with a finite spacing. 
With only one velocity measurement, \citet{Bricker_2007} proposed a method that linearly filters the \rev{energy} spectrum by assuming an inertial \rev{subrange} with a $-5/3$ slope.
Whereas \citet{Gerbi_2008} proposed fitting the measured spectrum to an empirical spectrum model to separate the waves and turbulence. 
While these methods are also frequently used due to their relative simplicity, they assume no wave-turbulence interactions and do not directly allow for a reconstructed time series (unless further approximations are made, e.g. see \citet{Cowherd_2021}). Additionally, some of the methods require multiple synchronized measurements which further restricts their use.

Some researchers sought a more rigorous characterization of the interaction of wave and turbulence by using the nonlinear wave stream  function \citep{Dean_1965}, which imposed fewer assumptions and allowed for a time reconstruction of the filtered signal.
\citet{Jiang_1990} developed a method based on this work that consists of solving, in a least squares sense, the stream functions at the free surface. 
Similarly, the triple decomposition method from \citet{Thais_1995, Thais_1996} also used a stream function to decompose the instantaneous velocity into turbulent fluctuation, irrotational wave, and rotational wave components.   
Although these methods allow the construction of separate wave and turbulent velocity time series, they are most well-suited to laboratory conditions where dispersive effects can be neglected \citep{Thais_1995}.

More recently, data-driven dimensionality-reduction techniques have been proposed to separate waves and turbulence. 
Ensemble Empirical Mode Decomposition (EEMD) is the foundation of the method developed by \citet{Huang_2008}.
It is based on the empirical mode decomposition (EMD), which decomposes the signal into intrinsic mode functions (IMFs), which define a basis dictated by the data \citep{Huang_1998}. While successful at separating a time series into different components, the method suffers because a single IMF can contain several oscillatory modes or a single mode can be split into several IMFs.
\citet{Huang_2008} proposed the EEMD to overcome this issue by introducing noise in the original signal in order to properly organize the different scales in the time series.
The limitations of this improvement are that a large number of iterations are needed to remove the effects of the noise and obtain an accurate decomposition. In addition, ensemble-averaging may affect the physical significance of the IMFs \rev{by artificially  redistributing some energy  among the IMFs}. 
More recently, \citet{Bian_2018} proposed the synchrosqueezed wavelet transform (SWT)-based method \citep{Daubechies_2011, Thakur_2013}, which decomposes the signal into a number of components separated in the time-frequency plane.  
This method forces the data to conform to a basis set determined \textit{a priori} and tends to underestimate the turbulence in the wave frequency range. 
The EEMD and SWT are modal decomposition techniques that can be applied to a single time series and allow for a time-resolved signal reconstruction. 
For these reasons, we use them to compare the outputs of our DMD-based method in section \ref{sec:ResultsAndDiscussion}. 

To summarize, traditional decomposition methods applied in spectral space are limited to scenarios with no wave-turbulence interactions, often require more specific assumptions and do not directly give a time-series reconstruction. 
To address these limitations, methods based on modeling waves using stream functions were developed, but they are only effective under ideal conditions that are not commonly found in nature. 
Finally, newer modal decomposition methods have been successful in generating a time-resolved filtered signal, but they can have a significant impact on the turbulence in the wave frequency range. 

\subsection{Modal decomposition of flow data}
\label{chp1:Sec_LitReview}
\rev{In this section, we present a summary of relevant modal decomposition algorithms; for a more in-depth reference, see \citet{Schmid_2022}.} Modal decomposition refers to the identification and separation of \emph{modes}, or pattern, in data from a dynamical system.
This type of decomposition is favorable because the general motion of a system, even nonlinear and nonstationary systems, can be approximated by a linear superposition of its modes. In the context of fluid dynamics, modal decomposition refers to the identification and separation of features in flow data (e.g., generated from Particle Image Velocimetry). 
The singular value decomposition (SVD) is the basis for the implementation of several dimensionality-reduction techniques such as proper orthogonal decomposition (POD) \citep{Lumley_1967} and dynamic mode decomposition (DMD) \citep{Schmid_2010}.

\subsubsection{Singular value decomposition (SVD) and proper orthogonal decomposition (POD)}
The SVD decomposes a matrix $\bm{\mathsf{M}} \in \mathbb{R}^{m \times n}$ into the product of three matrices
$$\bm{\mathsf{M}} = \bm{\mathsf{U}} \bm{\mathsf{\Sigma}} \bm{\mathsf{V}}^T,$$
where $\bm{\mathsf{U}} \in \mathbb{R}^{m \times m}$ and $\bm{\mathsf{V}} \in \mathbb{R}^{n \times n}$ are unitary matrices, and $\bm{\mathsf{\Sigma}} \in \mathbb{R}^{m \times n}$ is diagonal matrix with non-negative real values. The singular values that make up the diagonal values $\sigma_k \in \bm{\mathsf{\Sigma}}$ are in descending order and weight the modes. In other words, they are ordered based on how much of the total energy of the original data matrix they possess. The columns of $\bm{\mathsf{U}}$ (expressed as $\mathbf{u}_k$) and the columns of $\bm{\mathsf{V}}$ (expressed as $\mathbf{v}_k$) are known as the left singular vectors and right singular vectors of $\bm{\mathsf{M}}$, respectively. A reduced-order approximation of the matrix $\bm{\mathsf{M}}$ can be achieved by truncation as

$$\bm{\mathsf{M}}_r \approx \sum_{k=1}^r \sigma_k \mathbf{u}_k \mathbf{v}^T_k,$$
where $\bm{\mathsf{M}}_r$ is a rank $r$ approximation of $\bm{\mathsf{M}}$ using the $r \leq \min (m,n)$ most energetic singular modes. 

The POD is simply an SVD applied to a specifically constructed $\bm{\mathsf{M}}$ matrix. It was introduced in fluid mechanics by \citet{Lumley_1967, Lumley_2007}, \citet{Sirovich_1987, Sirovich_1987a, Sirovich_1987b}, and \citet{Aubry_1988}. 
The matrix $\bm{\mathsf{M}}$ is constructed as follows: given $n$ snapshots of flow data in a discrete spatial grid of $m$ points, each consecutive snapshot is reshaped into a corresponding column vector of a data matrix, $\bm{\mathsf{M}} \in \mathbb{R}^{m \times n}$, where the $k$-th column of $\bm{\mathsf{M}}$, denoted as $\mathbf{x}_k$, contains all the spatial data of a singular snapshot at time $k\Delta t$ or, similarly, each row contains a time series of the flow data at a single point in space \citep{Scherl_2020}. Thus we obtain:

$$\bm{\mathsf{M}} = \begin{bmatrix}
| & | & | & & |\\
\mathbf{x}_1 & \mathbf{x}_2 & \mathbf{x}_3 & ... & \mathbf{x}_n \\
| & | & |& & |
\end{bmatrix}.$$

\subsubsection{Dynamic mode decomposition (DMD)}
\label{subsubsec:DMD}

Dynamic Mode Decomposition \citep{Schmid_2010} is used to determine the temporal periodicity of coherent structures present in flow data. DMD captures the system dynamics in the eigenvectors (modes) and eigenvalues of an infinite-dimensional linear operator, known as the Koopman operator. 
DMD can separate system dynamics into modes that evolve over time. 
Additionally, DMD is tailored to recover oscillating dynamics without any prior knowledge of the system.

To compute DMD, one begins with $n$ time steps of data across $m$ dimensions. Each time step is shaped into a column vector of a present-state data matrix $\bm{\mathsf{X}} \in \mathbb{R}^{m \times (n-1)}$ that contains time steps $1$ to $n-1$, and a future-state data matrix $\bm{\mathsf{X}}' \in \mathbb{R}^{m \times (n-1)}$ that contains time steps $2$ to $n$. Thus we have
$$\bm{\mathsf{X}} = \begin{bmatrix}
| & | & | & & |\\
\mathbf{x}_1 & \mathbf{x}_2 & \mathbf{x}_3 & ... & \mathbf{x}_{n-1} \\
| & | & |& & |
\end{bmatrix}
\quad
\text{and}
\quad
\bm{\mathsf{X}}' = \begin{bmatrix}
| & | & | & & |\\
\mathbf{x}_2 & \mathbf{x}_3 & \mathbf{x}_4 & ... & \mathbf{x}_{n} \\
| & | & |& & |
\end{bmatrix}.
$$

Similar to POD, the $k$-th column of either $\bm{\mathsf{X}}$ or $\bm{\mathsf{X}}'$, denoted as $\mathbf{x}_k$, contains all the data of a single time step at time $k\Delta t$ or, similarly, each row contains a time series of the data for a single dimension.
The DMD algorithm attempts to find the best-fit linear operator $\bm{\mathsf{A}}:\bm{\mathsf{X}} \mapsto \bm{\mathsf{X}}' \in \mathbb{R}^{m \times m}$ that advances the current state of the system $\bm{\mathsf{X}}$ one time step into the future $\bm{\mathsf{X}}'$ as 
\begin{align}
\label{eqn:DMDsystem}
    \bm{\mathsf{X}}' \approx \bm{\mathsf{A}} \bm{\mathsf{X}}.
\end{align}
The solution to equation \ref{eqn:DMDsystem} \rev{can be approximated by a matrix $\bm{\mathsf{A}}$ that minimizes $\| \bm{\mathsf{X}}' - \bm{\mathsf{A}}\bm{\mathsf{X}} \|_F$ as}
\begin{align}
\label{eqn:DMDsystemSol}
    \bm{\mathsf{A}} = \bm{\mathsf{X}}' \bm{\mathsf{X}}^+,
\end{align}
where \rev{$\| \cdot \|_F$ is the Frobenius norm, and} $\bm{\mathsf{X}}^+$ is the pseudo-inverse of the matrix $\bm{\mathsf{X}}$.
Instead of directly finding the pseudo-inverse, the SVD of $\bm{\mathsf{X}}$ yields $\bm{\mathsf{X}} = \bm{\mathsf{U}} \bm{\mathsf{\Sigma}} \bm{\mathsf{V}}^T$ where $\bm{\mathsf{U}}$, $\bm{\mathsf{\Sigma}}$, and $\bm{\mathsf{V}}$ are unitary and therefore can be used to express the pseudo-inverse as $\bm{\mathsf{X}}^+ = \bm{\mathsf{V}} \bm{\mathsf{\Sigma}}^{-1} \bm{\mathsf{U}}^T$, which yields
\begin{align}
    \label{eqn:A}
    \bm{\mathsf{A}} = \bm{\mathsf{X}}' \bm{\mathsf{V}} \bm{\mathsf{\Sigma}}^{-1} \bm{\mathsf{U}}^T,
\end{align}
where the eigenvalues and eigenvectors (DMD modes) of $\bm{\mathsf{A}}$ capture the dynamics of the system. The eigenvalues of $\bm{\mathsf{A}}$ are complex numbers that capture the growth/decay and oscillation of its corresponding DMD modes in their real and imaginary components, respectively. The DMD modes of $\bm{\mathsf{A}}$ capture the spatial oscillatory patterns. 

Typically, calculating $\bm{\mathsf{A}}$ directly would be computationally intractable. Instead, a truncation of the matrices $\bm{\mathsf{U}}$, $\bm{\mathsf{\Sigma}}$, and $\bm{\mathsf{V}}$ to retain only the first $r$ singular values results in the reduced-rank matrices $\bm{\mathsf{U}}_r$, $\bm{\mathsf{\Sigma}}_r$, and $\bm{\mathsf{V}}_r$ and yields a best-fit linear operator $\Tilde{\bm{\mathsf{A}}} \in \mathbb{R}^{r \times r}$ by projecting $\bm{\mathsf{A}}$ onto the spatial modes $\bm{\mathsf{U}}_r$
\begin{align}
    \label{eqn:Atilde}
    \Tilde{\bm{\mathsf{A}}} = \bm{\mathsf{U}}_r^T \bm{\mathsf{A}} \bm{\mathsf{U}}_r = \bm{\mathsf{U}}_r^T \bm{\mathsf{X}}' \bm{\mathsf{V}}_r \bm{\mathsf{\Sigma}}_r^{-1}.
\end{align}

The operator $\Tilde{\bm{\mathsf{A}}}$ has the same discrete-time eigenvalues $\bm{\mathsf{\Lambda}}$ as $\bm{\mathsf{A}}$.
The eigenvectors $\bm{\mathsf{W}}$ of $\Tilde{\bm{\mathsf{A}}}$ can be calculated as $\Tilde{\bm{\mathsf{A}}}\bm{\mathsf{W}} = \bm{\mathsf{W}} \bm{\mathsf{\Lambda}}$ and converted to the high-dimensional eigenvalues $\bm{\mathsf{\Phi}}$ of $\bm{\mathsf{A}}$, as follows \citep{Tu_2014}:
\begin{align}
    \label{eqn:Aeigenvectors}
    \bm{\mathsf{\Phi}} = \bm{\mathsf{X}}' \bm{\mathsf{V}}_r \bm{\mathsf{\Sigma}}_r^{-1} \bm{\mathsf{W}}.
\end{align}

From the leading eigenvalues $\bm{\mathsf{\Lambda}}$ and eigenvectors $\bm{\mathsf{\Phi}}$ the solution of the system can be \rev{recreated} as
\begin{align}
    \label{eqn:sol2sys}
    \bm{\mathsf{X}} = \bm{\mathsf{\Phi}} \exp ({\bm{\mathsf{\Omega}} t}) \bm{\mathsf{b}} = \sum_{k=1}^r \phi_k \exp({\omega_k t}) b_k,
\end{align}
where $\bm{\mathsf{\Omega}} = \log (\bm{\mathsf{\Lambda}})/ \rev{\Delta}t$ is the continuous-time eigenvalue of $\bm{\mathsf{A}}$, and $\bm{\mathsf{b}} = \bm{\mathsf{\Phi}} \backslash \bm{\mathsf{X}}(t_0)$ is the initial condition of the system. 

\section{Methods}\label{sec:Methods}

\begin{figure}
   \centering
   \includegraphics[scale=.25]{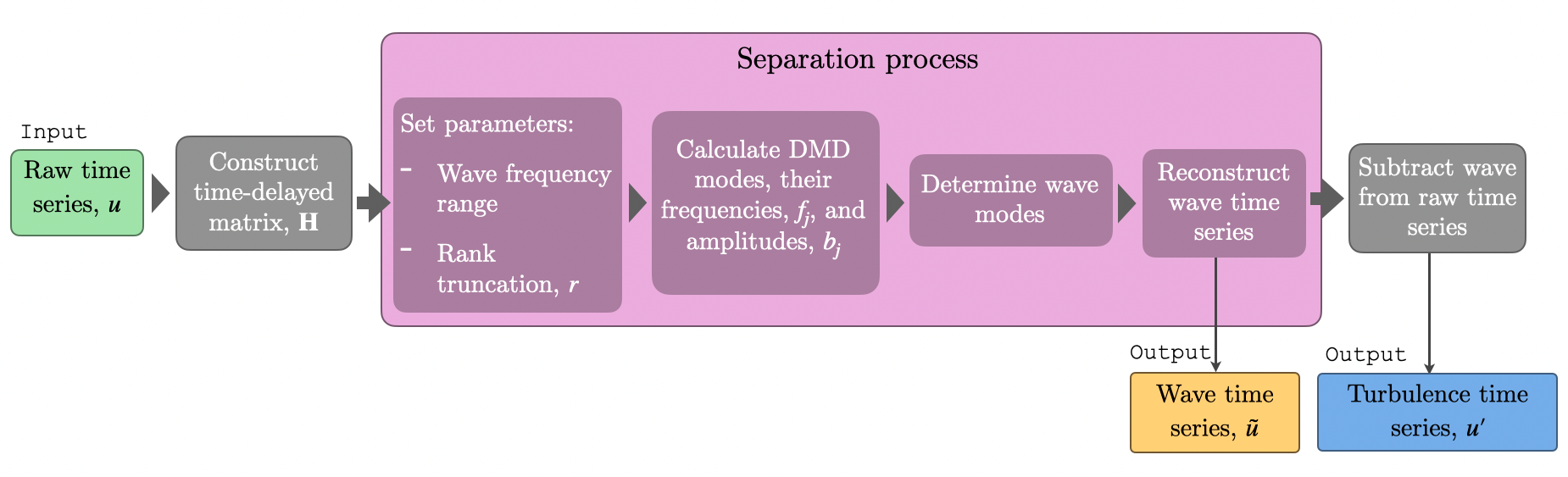}
   \caption{Workflow of the proposed DMD wave-turbulence separation methodology.}
   \label{fig:algorithmDiagram}
\end{figure}

Figure \ref{fig:algorithmDiagram} shows a diagram summarizing the algorithm workflow, which is described below.
For the method we propose, the input matrix contains a one-dimensional time series of discrete, point-wise flow measurements such as one velocity component ($u$, $v$, or $w$), or other variables like water elevation, scalar concentration, pressure, etc. 
These measurements are collected at a sampling frequency $f_s$ over a specified duration $T$, for $N=Tf_s$ total observations in the given time series.
Our approach is to use DMD to extract the most coherent features of the signal, which we assume to be the waves, as low-rank structures and subtract them from the raw signal to isolate the turbulence. 
Although other order-reduction techniques exist, they have limitations. 
For example, POD identifies correctly the rank of the system but since this method is based entirely on energy (or contained variance), the modes are not temporally orthogonal. 
In other words, POD can hierarchically rank coherent features in the signal based on their energy content but it cannot \rev{offer insights into how these patterns evolve over time}. 
This is not ideal for our problem because we expect waves to be represented as modes that have coherent oscillations in time. 
In contrast, DMD can separate the dynamics in the system and can identify their temporal oscillation frequency. This allows us to identify a specific frequency per mode that we can associate with wave motion \citep{Kutz_2016}.

In frequency space, waves can be narrow-banded when compared with turbulence, but their energy can still span over multiple orders of magnitude in frequency. The broader the wave frequency range, the larger the number of DMD modes will be needed to represent the waves. However, the number of modes that DMD can extract is limited by the rank of the input data matrix. Because we want to extract a range of wave modes, we need to increase the rank of our time series before we apply DMD.
 
Hence, we propose to address this problem and increase the rank of the input data matrix by constructing a delay embedding of the time-series signals, i.e.,  
\begin{equation}
\bm{\mathsf{H}} = \left[
\begin{array}{cccccc}
  x_{1} & x_{2} & x_{3} & \ldots & \ldots  &x_{n}  \\
   x_{2} & x_{3} &  &  & &\vdots \\
   x_{3} &  &  & & & \vdots \\ 
   \vdots & & & &  & x_{N-2}\\
   \vdots & &  & & x_{N-2}&  x_{N-1} \\
   x_{m} &  \ldots & \ldots & x_{N-2} & x_{N-1} & x_{N}
\end{array}  \right],
\label{eqn:Xdelayed}
\end{equation}
where $m = N - n$ is the number of time lags, and $n$ is the number of columns. Note that the matrix $\bm{\mathsf{H}}$ can have multiple rows besides the time-delay process if one has multiple data series (e.g., see \citet{Filho_2019, Fujii_2019, Lydon_2023}).
The introduction of the time delay serves a crucial role in increasing the rank of the data by increasing the ``information" content of the matrix \citep{Takens_1981}. 
Physically, this increase in rank can be analogous to Taylor's frozen turbulence hypothesis, since the time observations of the signals are converted to spatial observations.

In order to filter the signal using DMD, we first need to determine the wave frequency range, the dimension of the time-delayed matrix $\bm{\mathsf{H}}$, and the rank truncation $r$.
Visual inspection of the signal's power spectrum helps determine the wave frequency range by identifying the points where the wave peak meet the expected turbulence energy spectrum. 
For the construction of the time-delayed matrix, the number of columns $n$ needs to be large enough to capture the lowest wave frequencies of interest in each row. 
Based on our sensitivity analysis described below, we find that a good rule of thumb is within the range $n/N\approx [1/3, 1/2]$; however, in our data the decomposition is not as sensitive to this parameter as it is to the rank truncation $r$,  especially given a sufficiently long time series.
In general, we suspect that this parameter might require some tuning in order to determine the appropriate dimensions of $\bm{\mathsf{H}}$. 
From this new configuration $\bm{\mathsf{H}}$ of the input data, the present and future-state matrices $\bm{\mathsf{X}}$ and $\bm{\mathsf{X}}'$ are constructed as indicated in section \ref{subsubsec:DMD} as input for the DMD algorithm.
To determine the rank truncation $r$ we perform a POD decomposition of $\bm{\mathsf{H}} = \bm{\mathsf{U}}_{H}\bm{\mathsf{\Sigma}}_{H}\bm{\mathsf{V}}_{H}^T$ and compute the spectral density of each column $\mathbf{v}_j \in \bm{\mathsf{V}}_{H}$ (known as time expansion coefficients) as $\mathbf{C}_j = | \mathcal{F}\{\mathbf{v}_j\} |^2 / (f_s N)$\rev{, where $\mathcal{F}\{\cdot \}$ is the Fourier transform operator}. The vector $\mathbf{C}_j$ contains the energy content at each frequency of the $j$-th POD mode of $\bm{\mathsf{H}}$ \citep{Towne_2018}.
Plotting the energy content of each singular value mode $\mathbf{C}$  as a function of frequency helps us visualize which modes contain wave and/or turbulence energy. 
Next, we identify and discard the modes that do not contain wave energy. The parameter $r$ consists of the lowest modes that contain wave energy.
This truncation allows us to remove the higher-frequency turbulence from the data directly.

DMD is then applied to $\bm{\mathsf{H}}$, using the rank truncation $r$ that we just determined.  
Following the procedure outlined in the previous section, we obtain the eigenvalues $\bm{\mathsf{\Lambda}}$, eigenvectors (or modes) $\bm{\mathsf{\Phi}}$, and amplitudes $\bm{\mathsf{b}}$, which capture the coherent structures in the matrix $\bm{\mathsf{H}}$. 
The $j$-th DMD mode $\mathbf{\phi}_j$ and its corresponding eigenvalue $\lambda_j$ have a distinct oscillation frequency $f_j$ that can be calculated as
\begin{equation}
    f_j = \frac{\mathrm{Im}[\log (\lambda_j)]}{2\pi \rev{\Delta}t}.
\end{equation}
The modes with frequencies within the wave range are referred to as wave modes and are transformed to the time domain as a wave time-series using equation \ref{eqn:sol2sys}.
The wave time-series is then subtracted from the raw signal, leaving the turbulence component of the signal. 

\section{Dataset descriptions}
\label{sec:DatasetDescription}

\label{sec:Dataset}
We test our method on three distinct datasets to show its broad applicability: a synthetic dataset with known wave and turbulence components; a field dataset collected on a swell-dominated coastal shelf in a bottom boundary layer published in \citet{Reimers_2021}; and laboratory data we collected in a surface boundary layer in a wind-driven wave tank. Table \ref{tab:summaryDMDparams} shows some flow and DMD parameters used in our proposed decomposition method for each dataset. Additionally, it shows the relative wave intensity $\tilde{u}_{\text{rms}}/u'_{\text{rms}}$, which is estimated as the ratio of the wave to the turbulence root-mean-square velocity obtained from the DMD-based decomposition. Note that we cover a range of wave intensities and that the field and laboratory waves are the most and least intense, respectively. 

\begin{table}[htb]
    \centering
    \begin{tabular}{cccccccccccc}
        \topline
         Data type & $N$ & $f_s$ (Hz) & $T_p$ (s) & Wave range (Hz) & $m$ & $n$ & $r$ & $\psi_{\text{wave}}$ & $r/n$ & $\psi_{\text{wave}}/r$ & $\tilde{u}_{\text{rms}}/u'_{\text{rms}}$\\
         \midline
        Synthetic & 10000 & 10 & 5 & $0.117 - 0.6~~~$ & 3000 & 5911 & 173 & 104 & 0.029 & 0.601 & 1.91 \\
        Field & 7200 & 8 & 14.29 & $0.031 - 0.175$ & 5700 & 1500 & 48 & 34 & 0.027 & 0.708 & 8.05 \\
        Laboratory, $u$ & 18000 & 30 & 0.5 & $1.02 - 3.16$ & 13000 & 5000 & 1000 & 506 & 0.200 & 0.506 & 0.30\\
        Laboratory, $w$ & 18000 & 30 & 0.5 & $1.02 - 3.16$ & 12500 & 6500 & 1300 & 732 & 0.200 & 0.563 & 0.72\\
        \botline
    \end{tabular}
    \caption{Summary of dataset characteristics including some input and output parameters from the wave-turbulence decomposition. Characteristics include length of time series $N$, sampling frequency $f_s$, wave peak period $T_p$, and wave frequency range. We also report the optimal number of rows $m$ and columns $n$ of the time-delayed matrix $\bm{\mathsf{H}}$, the optimal rank truncation $r$ used and the number of wave DMD modes $\psi_{\text{wave}}$, including the relative rank $r/n$  used to separate high frequency turbulence unaffected by the wave and $\psi_{\text{wave}}/r$ used to reconstruct the wave time series. The final column reports the relative wave intensity $\tilde{u}_{\text{rms}}/u'_{\text{rms}}$ using the DMD-separated velocities. \rev{For the synthetic data, the true wave intensity is $\tilde{u}_{\text{rms}}/u'_{\text{rms}} = 1.81$.}}
    \label{tab:summaryDMDparams}
\end{table}

\subsection{Synthetic data}
Synthetic data allows us to test our wave-turbulence decomposition by comparing our decomposed velocities to the ground-truth signals directly. 
To create the dataset, we linearly add a wave $\tilde{u}$ and turbulence $u'$ velocity time series. In this case, wave-turbulence interactions are exactly zero and $\tau/\rho = \rev{-}\langle u'w'\rangle$.  
The turbulence data come from a streamwise velocity time series collected with hot-wire anemometry in a wind tunnel sampled at 60 kHz \citep{Castro_2021}. 
The turbulence spectrum has a developed inertial range between 0.13 Hz to 0.6 Hz. 

The wave amplitude spectrum $S(\omega)$ was generated using a Joint North Sea Wave Project (JONSWAP) model \citep{Hasselmann_1973} 
\begin{equation}
    S(\omega) = \frac{\alpha g^2}{\omega^5}\exp \left [ -\frac{5}{4} \left ( \frac{\omega_p}{\omega}\right ) \right ] \gamma^{\exp\left [ -\frac{(\omega/\omega_p - 1)^2}{2\beta^2} \right ]},
\end{equation}
where $\alpha = 0.01$ is the intensity of the JONSWAP spectrum, $g$ is the acceleration of gravity, $\omega$ is the angular frequency, $\omega_p$ is the angular frequency of the peak of the spectrum, $\gamma = 3.3$, and the factor $\beta = 0.07$ for $\omega \leq \omega_p$ and $\beta = 0.09$ for $\omega > \omega_p$. 
The peak period $T_p$ was chosen to be $5$ seconds, which results in $\omega_p = 0.2$ Hz. At a point below the surface where $x=0$, and $z= -2$ m,  linear wave theory was used to generate the velocity time series from the sum of $N_w=500$ wave components:
\begin{align}
    \tilde{u} (x, z, t) &= \sum_{i=1}^{N_w} U_i \cos(k_ix - \omega_it + \phi_i)\exp (k_i z),
\end{align}
where $U_i = a_i \omega_i$ is the wave velocity amplitude vector, $a_i = \sqrt{2 S(\omega_i) d\omega}$ is the amplitude of the spectrum at frequency $\omega_i$, $k_i = \omega_i^2/g$ is the wave number that satisfies the deep-water linear dispersion relation, and $\phi_i$ is the phase which is randomly sampled from the interval $[0,\, 2\pi)$.

The turbulence time series was decimated from 60 kHz to 10 Hz to match the sampling frequency of the JONSWAP waves time series.
The total detrended horizontal component of velocity $u$ is obtained from adding the wave and turbulence components of horizontal velocity, defined in equation \ref{eq:decomp}, as $u = u' + \tilde{u}$.
The power spectrum of $u$, shown in Fig. \ref{fig:SynthDataDMDdecomp}b, preserves the turbulence developed inertial range between 0.13 Hz to 0.6 Hz and shows that the wave energy is concentrated around the wave peak frequency at 0.2 Hz.

\subsection{Field data}
The field data consists of long waves measured in a turbulent boundary layer above the coastal shelf.
The data comes from a published dataset in \citet{Reimers_2021}. 
We are analyzing their acoustic Doppler velocimeter (ADV) data collected during a campaign to measure oxygen eddy covariance. 
The ADV captures a time series of the three perpendicular velocity components and pressure at a single point in space. 
We analyze only a time series of $u$ velocity components from their data. 
The measurements were collected from a lander deployment along the Oregon shelf above a sandy seafloor.  
The ADV probe's sampling volume was 30 cm above the bottom of the sea floor in 80 m of water and was sampling at a frequency of $f_s = 64$ Hz, which was then downsampled to $f_s = 8$ Hz. 
Each deployment was broken down into 15-min (7200 observations) long bursts, and we consider only one burst in our study.
The mean current of the analyzed burst time series is $0.11$ m/s,  the peak wave period $T_p$ is $14.28$ seconds (see Fig. \ref{fig:FieldDataDMDdecomp}b) and for reference, the significant wave height was estimated to be 3.4 m. 
We chose this specific burst because it presented a mix of high and low amplitude waves, as shown in Fig. \ref{fig:FieldDataDMDdecomp}c. 
For further details about the data acquisition, please consult the original study \citep{Reimers_2021}.

\subsection{Laboratory data}
To contrast the bottom boundary layer field data, we collected laboratory data in a wind-driven surface boundary layer with short intermittently breaking waves. 
The data was collected during experiments conducted in the Washington Air-Sea Interaction Facility (WASIRF) which is a 12.2 m long, 0.91 m wide wind-wave tank. 
Fresh water was filled to a depth of 0.6 m leaving 0.6 m of headspace above for air circulation. 
To create wind, a suction fan propelled air through the test section's headspace and recirculated it via an overhead duct. See \citet{Baker_2023} for more details on the facility. 

We collected a timeseries of 3-dimensional velocity at a height of 19 cm below the water surface with an ADV (Nortek Vectrino) under a wind speed of 16 m/s at a fetch of 7.5 m.
The data was collected over 20 minutes at a sampling frequency $f_s = 30$ Hz. The wave energy spans the frequency range of 1 to 3 Hz, with a peak at $2$ Hz (Figs. \ref{fig:LabDataDMDdecomp_u}b and \ref{fig:LabDataDMDdecomp_w}b).
We post-processed the data using the despiking method proposed by \citet{Goring_2002}. 
Even after despiking the data, there is still some noise present in the data due to intermittent wave-breaking which generated bubbles which affected the data quality. 
This noisy data provides another challenge to the decomposition.
We apply our decomposition method to both the horizontal and vertical components of velocity because they present contrasting conditions. 
The horizontal velocity data has high turbulence fluctuations when compared to the wave velocities, whereas the vertical velocity has wave and turbulence fluctuations with similar magnitude (see Table \ref{tab:summaryDMDparams}). 
Finally, we note that due the strong anisotropy in the surface boundary layer flow, we see deviation from the -5/3 slope in the inertial range \rev{(Fig. \ref{fig:LabDataDMDdecomp_u}b)}.

\section{Results and Discussion}
\label{sec:ResultsAndDiscussion}
In this section, we present the results obtained from analysing the synthetic, laboratory, and field data with our proposed DMD method.
We demonstrate our DMD decomposition by applying it to all three datasets after first conducting a sensitivity analysis on the synthetic data. 
Next, we compare our results to the other two time-resolved mode-based decomposition methods: EEMD \citep{Qiao_2016, Huang_2009} and Synchrosqueezed Wavelet transform (SWT) \citep{Bian_2018}. 
After  assessing the decomposition in both spectral and time space, we analyze the temporal autocorrelation of the decomposed turbulence signal, further demonstrating how our DMD method outperforms the other methods. 

\subsection{Synthetic data decomposition}

\begin{figure}[h]
    \centering
    \includegraphics[scale=1]{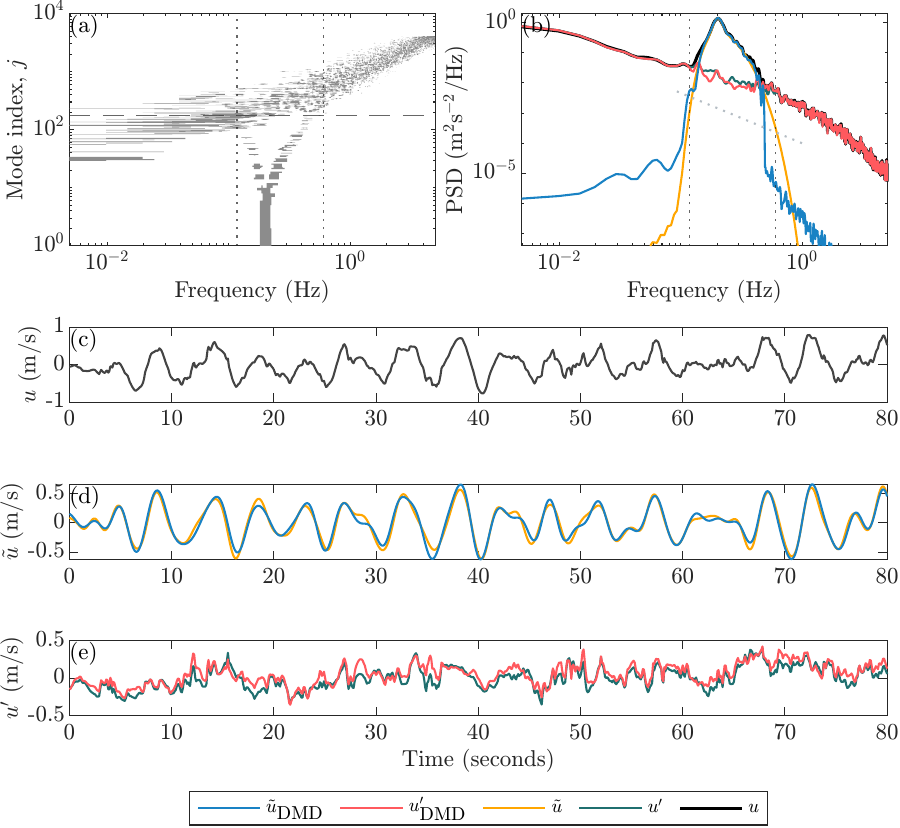}
    \caption{DMD wave-turbulence decomposition of $u$-component velocity of synthetic data. (a) POD spectrum of time-delayed matrix $\bm{\mathsf{H}}$. (b) Power spectra of the original data and DMD-separated wave and turbulence. A -5/3 slope line is plotted for reference. In (c) we plot a portion of the original time series, (d) the true and separated wave components using 104 DMD modes, (e) and the true and separated turbulence components.}
    \label{fig:SynthDataDMDdecomp}
\end{figure}

The first step of our decomposition method is constructing the time-delayed matrix $\bm{\mathsf{H}}$ and identifying the wave frequency range, as described in section \ref{sec:Methods} and shown in Table \ref{tab:summaryDMDparams}. 
Next, we determine the rank truncation $r$ by plotting the POD spectrum of $\bm{\mathsf{H}}$ of the synthetic $u$ data (Fig. \ref{fig:SynthDataDMDdecomp}a) and selecting the modes that contain wave energy.  
We observe two strong features in this plot: (1) the  higher modes contain energy in monotonically increasing frequencies across the entire domain due to turbulence, showing the energy \rev{decreasing from low to high frequencies consistent with the inertial range} and (2) a coherent high energy signal in lower modes between the frequency range of 0.13 and 0.6 Hz indicated by the vertical dotted lines.
This ``dip'' in the POD spectrum indicates an energetic range associated with the wave energy and accounts for a significant part of the variance in the time series.
We use the location of the dip to inform the rank truncation. 
As indicated with the horizontal dashed line, we truncate at the $r=173$ mode; that is, we take only the first 173 modes for the next step in the DMD. 
We see that while most of the wave energy is contained in these first $r$ modes, there is also energy associated with low-frequency turbulence in these lowest modes. 
This clearly shows how POD alone fails at separating waves and turbulence since it can only isolate the high-frequency turbulence from the waves, not the lower-frequency turbulence. 

After determining the rank truncation, we are able to decompose the signal using DMD. Figure \ref{fig:SynthDataDMDdecomp}b shows the spectra of the DMD-filtered wave and turbulence time series which we compare with the true wave and true turbulence spectra. Within the wave-frequency range, the DMD-filtered wave spectrum $\tilde{u}_{\textrm{DMD}}$ closely follows the wave peak. Outside the wave-frequency range, the $\tilde{u}_{\textrm{DMD}}$ spectrum falls sharply, capturing a similar shape to the true wave spectrum for multiple orders of magnitude before levelling off. The mismatch in the signals is clearer to see in the $u'_{\textrm{DMD}}$ spectrum, which slightly underpredicts the turbulence in the wave-frequency range.

Figures \ref{fig:SynthDataDMDdecomp}(c-e) show the full $u$ time series and the decomposed $\tilde{u}$ and $u'$ time series alongside their true counterparts. We see overall agreement between the decomposed and true time series, however some mismatch exists.
While the turbulence time-series plot shows that the separated signal accurately captures the low frequency turbulence fluctuations, there is stronger disagreement in the fluctuations with frequencies in the wave-range. 
Given that the wave signal was generated with 500 discrete wave signals and that the DMD only resolved 104 modes, we do not expect a perfect reconstruction. 
In this method, the DMD is discretizing a wave frequency range that in reality is continuous, so we always expect some mismatch.
Column $r/n$ in Table \ref{tab:summaryDMDparams} shows the ratio of the number of modes used to truncate the input matrix normalized by the number of columns used to construct the input matrix. 
For the synthetic data, the low rank-to-column ratio indicates that the dominant wave patterns in the signal were reconstructed using fewer modes compared to the other datasets.

\begin{figure}[h]
    \centering
    \includegraphics[scale=1]{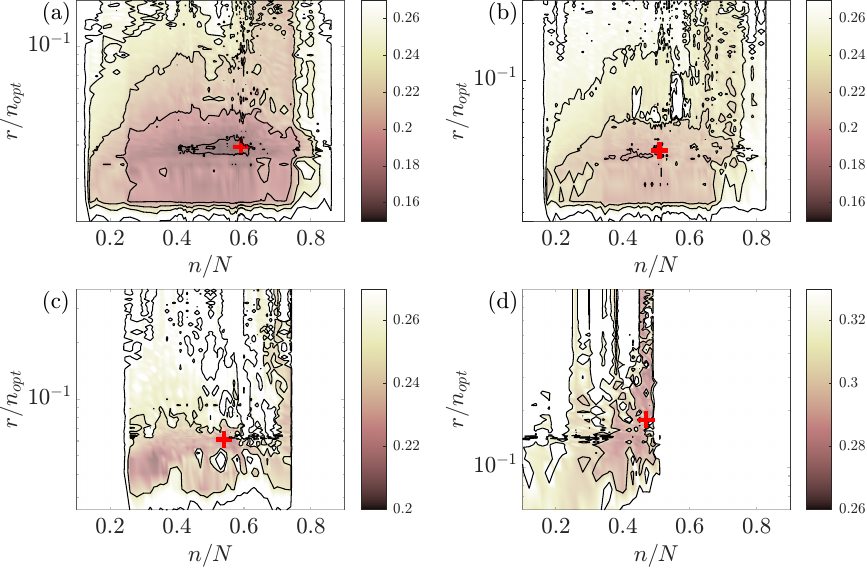}
    \caption{Sensitivity analysis to asses the separation performance of the DMD-based algorithm for different time series length, $N$. A measurement of error, $\epsilon$, was calculated for every combination of normalized rank truncation, $r/n_{\text{opt}}$, and normalized number of columns, $n/N$, in the time-delayed matrix, $\bm{\mathsf{H}}$, where $n_{\text{opt}}$ is the optimal number of columns and the optimal combination is indicated by a red cross. (a) $\epsilon = 0.1586$, at $r_{\text{opt}}/n_{\text{opt}}=0.029$, $n_{\text{opt}}/N = 0.5911$ ($N= 10000$), (b) $\epsilon = 0.1894$, at $r_{\text{opt}}/n_{\text{opt}}=0.043$, $n_{\text{opt}}/N = 0.509$ ($N=7500$), (c) $\epsilon = 0.2109$, at $r_{\text{opt}}/n_{\text{opt}}=0.061$, $n_{\text{opt}}/N = 0.540$ ($N=5000$), and (d) $\epsilon = 0.2737$, at $r_{\text{opt}}/n_{\text{opt}}=0.179$, $n_{\text{opt}}/N = 0.470$ ($N=2500$).}
    \label{fig:SynthDataSensAnalysis}
\end{figure}

\begin{table}
\begin{subtable}[c]{0.5\textwidth}
\centering
\begin{tabular}{ c c c c c c }
        \topline 
      \multirow{2}{*}{\quad Parameters \quad} & \multicolumn{4}{c}{Length of raw time series, $N$} \\
      \cline{2-5}
      \multirow{2}{*}{} & \quad 2500 \quad & \quad 5000 \quad & \quad 7500 \quad & \quad 10000 \quad \\
      \midline
      $r_{\text{opt}}$ & 210 & 165 & 155 & 173 \\
      $r_{\text{opt}}/n_{\text{opt}}$ & 0.179 & 0.061 & 0.043 & 0.029 \\
      $n_{\text{opt}}$ & 1175 & 2700 & 3900 & 5911\\
      $n_{\text{opt}}/N$ & 0.47 & 0.54 & 0.52 & 0.59\\
      \hline
      $\epsilon$ & 0.2737 & 0.2109 & 0.1894 & 0.1586\\
      \botline
  \end{tabular}
\label{tab:SensAnalysisNumObs}
\end{subtable}%
\begin{subtable}[c]{0.5\textwidth}
\centering
\begin{tabular}{cccc}
        \topline
         Method & $\quad \epsilon \quad$ & $\quad \alpha$ (m/s) \quad & $\quad \alpha / u'_{\text{rms}}\quad$\\
         \midline
        DMD & \quad 0.159 \quad & \quad 0.039 \quad & \quad 0.24 \quad \\
        EEMD & \quad 0.268 \quad & \quad 0.065 \quad & \quad 0.40 \quad \\
        SWT & \quad 0.325 \quad & \quad 0.079 \quad & \quad 0.49 \quad \\
        \botline
    \end{tabular}
\label{tab:summaryErrorAnalysis}
\end{subtable}
\caption{(Left) Summary of sensitivity analysis from Fig. \ref{fig:SynthDataSensAnalysis} and (Right) comparison of errors from DMD and other modal decomposition methods applied to the longest time-series length of $N=10000$.}
\label{tab:summaryDMDsensityvityAnalaysis}
\end{table}

\subsubsection{Sensitivity analysis on decomposition parameters}
We conducted a sensitivity analysis on the decomposition based on the two main tuning parameters: the rank truncation $r$ and the number of columns of the time-delayed input matrix $n$ in order to identify the optimal $r_{\text{opt}}$ and $n_{\text{opt}}$ for the decomposition. In addition, we also varied the length of the time series $N$ and repeated the sensitivity analysis to assess how $r$ and $n$ vary with $N$. 
We quantified the accuracy of the decomposition by minimizing the normalized error $\epsilon$, which is defined as follows:
\begin{equation}
    \epsilon = \frac{\| u'_i-u'_{i,\text{DMD}}\|_2}{\|u'_i\|_2},
\end{equation}
where $u'_i$ and $u'_{i,\text{DMD}}$ are the true and decomposed turbulence time series (here the notation refers to DMD, but we also calculate this error for the EEMD and SWT methods), respectively, at time $t_i$, and $\| \cdot \|_2$ is the $L_2$ norm. For comparison, we also measured the mean absolute error $\alpha$ defined as
\begin{equation}
    \alpha = \frac{1}{N} \sum_{i=0}^N |(u'_i-u'_{i,\text{DMD}})|.
\end{equation}
 
We conduct the sensitivity analysis across four different time-series lengths to assess the importance of the length of the dataset.
Figures \ref{fig:SynthDataSensAnalysis}(a-d) show the results of the sensitivity analysis for $N=10000, 7500, 5000$, and $2500$, respectively.   We plot the norm-based error $\epsilon$ for different combinations from the space of normalized $r/n_{\text{opt}}$ and normalized number of columns $n/N$. The optimal combination $(n_{\text{opt}}/N,\, r_{\text{opt}}/n_{\text{opt}})$ is indicated by a red cross.

Overall, the optimal $n/N$ ratio is not too sensitive and ranges between 0.5 and 0.6, while the optimal $r$ minimizes the error only in a narrow range of values.
This indicates that, at least for this simple case, the preferential dimension of the input time-delayed matrix is an approximately square matrix. This might be due to the fact that for a large dataset we have enough physical information to resolve coherent features in the time domain $n$ and the ``spatial" domain $m$ that we constructed via time-delaying. However, when there is limited information by the shortened time-series, as in the case of Fig. \ref{fig:SynthDataSensAnalysis}d, the preferential input matrix shape is a tall, skinny matrix. This suggests a preference for ``spatial" information over temporal information, and that meeting a critical column length is more important than increased row length. 
We can observe that the optimal $r$ shows a small decrease with $N$ (except for $N=2500$) but overall is fairly stable.  
Selecting an appropriate rank truncation is essential for accurately capturing the waves in the time series. If $r$ is too low, some waves may be missed by the DMD algorithm, leading to an incomplete reconstructed time series. Conversely, if $r$ is too high, it may lead to overestimation of the wave energy. 

Table \ref{tab:summaryDMDsensityvityAnalaysis} summarizes the different tuning parameters for all time-series length $N$ and their associated errors. We see a decrease of the norm-based error $\epsilon$ with $N$ which makes sense since the DMD is a data-based algorithm, and more data generally improves performance. However, as the data increases so does the computational cost of performing the decomposition. We compare the DMD method to EEMD and SWT in Table \ref{tab:summaryDMDsensityvityAnalaysis}, where we see that given the same time-series length, our DMD method outperforms the other modal decomposition methods. 
While the EEMD outperforms the SWT, they both  have higher errors than the DMD method, even when DMD was applied to only half the data ($N=5000$). Overall, our method seems to perform the best when applied to this dataset. At the end of this section, we compare the methods in more detail by applying them to all of the datasets.

\rev{In our analysis, we use the ratio $r/n$ to understand how effectively the first step of our algorithm, the rank truncation, separates waves from turbulence. A low $r/n$ indicates a more effective rank truncation.
This ratio measures how well we can isolate the wave energy from the higher frequency turbulence that does not overlap in the frequency space. 
Essentially, this step resembles a low-pass filter, in the sense that most of the truncated modes correspond to high frequency motion.  
We don’t expect $r/n$ to be universal because the shape of the spectra can vary; e.g., narrow-banded waves at lower frequencies would likely result in a lower $r/n$ relative to  wide-banded waves with the same intensity that occupy higher frequencies, given the same turbulence spectrum. 
In addition, the number of columns $n$ in the input matrix controls the lowest resolvable frequencies, and therefore will depend on the location of the wave frequency peak in frequency domain; higher-frequency waves might require a smaller $n$ compared to low-frequency waves  in order to capture the peak wave frequency.
Our observations show that $r/n$ generally increases as wave intensity decreases, indicating that the rank truncation is less effective at isolating wave energy when the waves are weaker.}

\rev{We next consider $\psi_{\text{wave}}/r$ to evaluate the effectiveness of the second step of our decomposition (see Table \ref{tab:summaryDMDparams}). 
This ratio represents the proportion of DMD modes used to reconstruct the waves relative to all modes remaining after the initial truncation. 
Higher wave intensity usually corresponds to a higher $\psi_{\text{wave}}/r$ because most of the modes retained are wave modes. 
Essentially, to summarize the two main steps of the algorithm: first the rank truncation initially identifies the modes containing the relevant frequency content, and second the DMD step extracts the wave components from this subset. 
By quantifying  $r/n$ and $\psi_{wave}/r$, we can compare the effectiveness of each step in the separation process.}

\begin{figure}[htbp]
        \centering
        \includegraphics[scale=1]{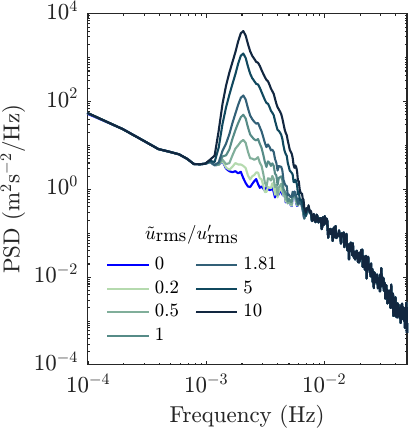} 

    \caption{Synthetic spectra used for sensitivity analysis on wave intensity $\Tilde{u}_{\text{rms}}/u'_{\text{rms}}$. Wave intensity of zero represents the raw turbulence signal, $u'$.}
    \label{fig:Synth_waveInt_spec}
\end{figure}

\subsubsection{Sensitivity analysis on wave intensity}
\begin{figure}[htpb]
    \centering
    \includegraphics[scale=1]{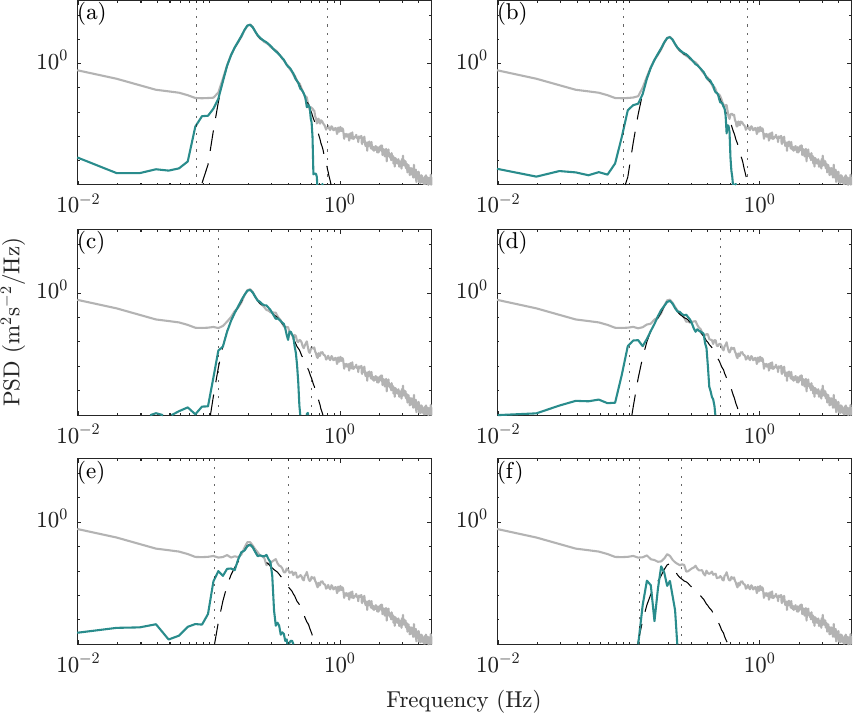}
    \caption{Comparison of the DMD-separated wave spectrum (solid teal line) to the true wave spectrum (dashed black line) and the combined wave and turbulence spectrum (solid grey line). Each subfigure correspond to a wave intensity: (a) $\Tilde{u}_{\text{rms}}/u'_{\text{rms}} = 10$; (b) $\Tilde{u}_{\text{rms}}/u'_{\text{rms}} = 5$; (c) $\Tilde{u}_{\text{rms}}/u'_{\text{rms}} = 1.81$; (d) $\Tilde{u}_{\text{rms}}/u'_{\text{rms}} = 1$; (e) $\Tilde{u}_{\text{rms}}/u'_{\text{rms}} = 0.5$; (f) $\Tilde{u}_{\text{rms}}/u'_{\text{rms}} = 0.2$.} 
    \label{fig:Synth_waveInt_waveSpec_vs_trueWaveSpec}
\end{figure}

\rev{Additionally, we performed a sensitivity analysis on the effect of relative wave intensity $\Tilde{u}_{\text{rms}}/u'_{\text{rms}}$. We scaled the synthetic wave time series amplitudes linearly to adjust the wave intensity, keeping the frequency and phase components of the synthetic wave signal constant. 
This results in the wave spectrum moving up or down relative to the turbulence spectrum, as shown in Fig. \ref{fig:Synth_waveInt_waveSpec_vs_trueWaveSpec}. 
This also changes the apparent width of the wave spectral peak which overlaps the inertial range; a change that is due to the weaker wave energy away from the peak frequency being overshadowed by the turbulence.
We chose the wave intensities $\Tilde{u}_{\text{rms}}/u'_{\text{rms}} = [0.2, 0.5, 1, 5, 10]$, in addition to the wave intensity $\Tilde{u}_{\text{rms}}/u'_{\text{rms}} = 1.81$ that we used in the previous single-case analysis (Fig. \ref{fig:Synth_waveInt_waveSpec_vs_trueWaveSpec}), to cover a wide range of scenarios. 
This range also covers the high and low intensities present in the field and laboratory data, respectively.}

\begin{figure}[htpb]
        \centering
        \includegraphics[scale=1]{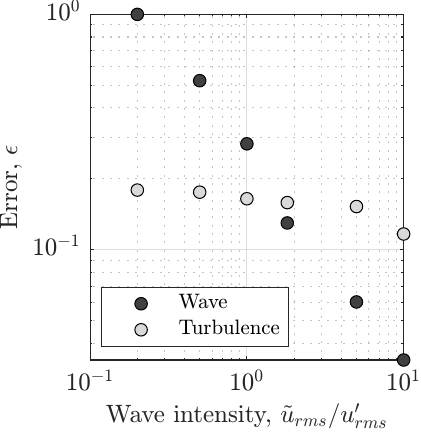}

    \caption{Reconstruction error from wave and turbulence at the different wave intensities in the sensitivity analysis.}
    \label{fig:Synth_waveInt_Error}
\end{figure}

\rev{Figure \ref{fig:Synth_waveInt_Error} presents the norm-based reconstruction errors for waves and turbulence across varying wave intensities.  
We expect a decrease in the reconstruction error as wave intensity increases because the DMD algorithm will be able to more effectively identify the coherent features of wave motion as distinct from the turbulence. 
Conversely, as wave intensity decreases, the performance of our decomposition method is expected to decline. 
We note that the error in turbulence reconstruction decreases more slowly with increasing wave intensity, relative to the wave reconstruction error. 
This is because the turbulence reconstruction is less sensitive to the decomposition because most of the turbulence in the signal is unaffected by the waves due to it occurring at frequencies outside the wave range.}

\rev{In Fig. \ref{fig:Synth_waveInt_waveSpec_vs_trueWaveSpec}, we plot the decomposed wave spectra relative to the input spectra and the true wave spectra to further evaluate the effectiveness of the decomposition. 
We see that in all cases, the wave peak above the turbulence spectra is well represented by the decomposition. However, as the wave intensity decreases and more of the wave spectrum falls below the turbulence spectrum, we see the decomposition struggles to recover the shape of the original wave spectrum accurately. 
This is especially clear in the lowest wave intensity case, which has a large wave reconstruction error due to the poor reconstruction of the wave and the small norm of the true wave signal. 
Additionally, we see that the DMD method tends to overestimate the wave energy at the lower end of the wave frequency range, while underestimating the wave energy at the higher end. 
Although, these errors are small compared to the overall spectral energy in the combined wave and turbulence signal. 
In summary, the DMD method works best when the wave intensity is high and fails when the wave intensity is low; therefore, we recommend our method for cases when the wave energy is similar or larger in magnitude than that of the turbulence  and for when there exists a clear wave peak in the power spectrum of the raw signal.}

\begin{figure}[htpb]
        \centering
        \includegraphics[scale=1]{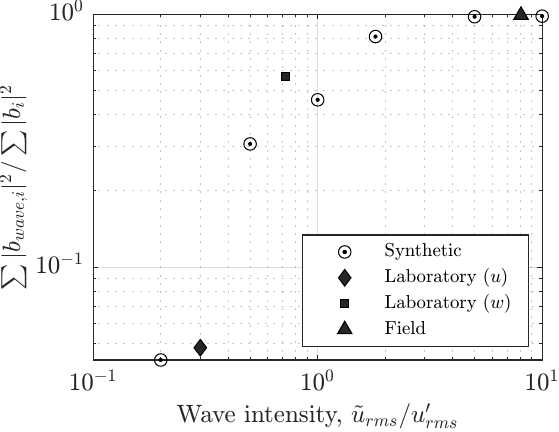}

    \caption{Relative energy contained in the DMD modes that each wave intensity case requires to resolve the coherent features in the signal. The $y$ axis is the energy of the DMD wave modes relative to the energy contained in the entire signal, where $b_{i}$ represents the amplitude of the $i$-th DMD mode, while $b_{\text{wave},i}$ represents the amplitude of the subset of DMD modes associated with wave motion and used in the reconstruction of the wave signal.}
    \label{fig:Synth_waveInt_waveInt_vs_relEnergy_log}
\end{figure}

\rev{To further evaluate how the decomposition performs across wave intensity, we consider the relative energy of the DMD wave modes. 
Figure \ref{fig:Synth_waveInt_waveInt_vs_relEnergy_log} shows a direct correlation between the amount of wave energy in the signal and the corresponding DMD mode energy required to accurately reconstruct the wave signal.
The expression $ \sum |b_{\text{wave}, i}|^2 / \sum |b_{i}|^2$ in the $y$ axis in Fig. \ref{fig:Synth_waveInt_waveInt_vs_relEnergy_log} quantifies the ratio of the energy contributed by the subset of wave-containing DMD modes to the total energy from all DMD modes.
Values close to unity mean that the wave modes dominate the dynamics of the system, while values close to zero mean that the contribution of the wave modes is small compared to the overall dynamics.
Figure \ref{fig:Synth_waveInt_waveInt_vs_relEnergy_log} also shows a general trend across data sets and wave intensities.
When waves and turbulence are similar in intensity ($\Tilde{u}_{\text{rms}}/u'_{\text{rms}} = 1$), the waves account for approximately half of the energy in the DMD modes.
However, as the wave intensity nearly doubles ($\Tilde{u}_{\text{rms}}/u'_{\text{rms}} = 1.81$), the proportion of DMD mode energy associated with waves increases significantly, indicating a strong dominance of wave energy in the reconstruction. Conversely, when the wave intensity is halved, waves contribute only about 30$\%$ of the DMD mode energy, highlighting the reduced influence of weaker wave signals.}

\subsection{Field data decomposition}

\begin{figure}[htbp]
    \centering
    \includegraphics[scale=1]{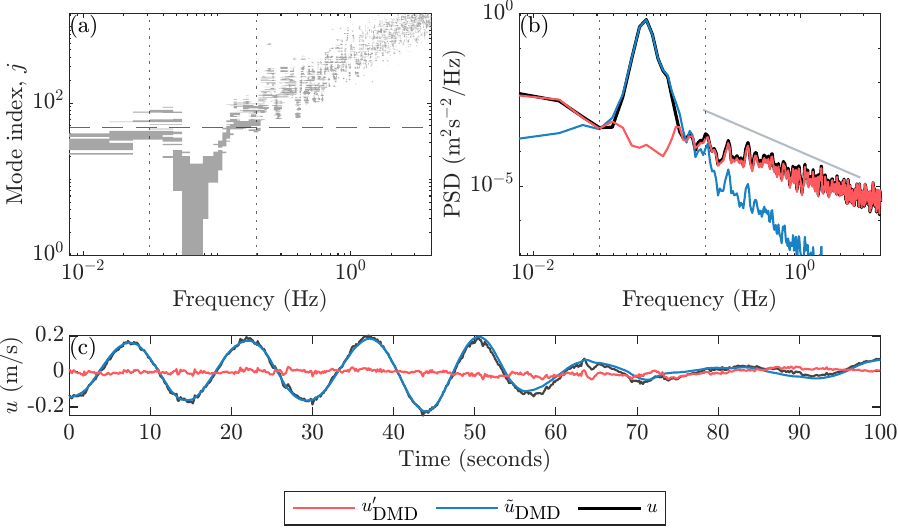}
    \caption{DMD wave-turbulence decomposition of $u$-component velocity of field data. (a) POD spectrum of time-delayed matrix $\bm{\mathsf{H}}$. (b) Power spectra of the original data and DMD-separated wave and turbulence. \rev{A gray -5/3 slope line is plotted for reference.} (c) A portion of the time-series of the original and decomposed signals. The wave motion was reconstructed using 34 DMD modes.}
    \label{fig:FieldDataDMDdecomp}
\end{figure}

We plot the POD spectrum of the field data (Fig. \ref{fig:FieldDataDMDdecomp}a) and find that, similar to the synthetic data in Fig. \ref{fig:SynthDataDMDdecomp}a, there are two features: one associated with the turbulence energy that is seen across the frequency range and one that marks a dip in the energetic modes associated with the the wave energy concentrated between 0.03 and 0.2 Hz, demarcated with the vertical dotted lines.
\rev{It is worth noting that in this case, there is not a well-defined turbulence signal in the wave frequency range. 
Wave and turbulence interactions present in real data, as compared with the synthetic data, may reduce the effectiveness of the POD modal separation in the overlapping frequency range. 
This further supports why POD alone cannot always isolate waves and turbulence.}
Based off the location of the dip, we rank-truncate to $r=48$ modes to preserve the energy in the wave frequencies and remove the high frequency turbulence.
The final filtered velocity spectra are shown in  Fig. \ref{fig:SynthDataDMDdecomp}b alongside the raw signal spectrum. 
We see that the DMD method is able to isolate the wave energy and flatten the turbulence spectrum under the wave peak. While the wave spectrum contains nonzero energy in the frequencies outside the wave range, the energy is small enough to not affect the turbulence spectrum. 

In Fig. \ref{fig:SynthDataDMDdecomp}c, we present a portion of the separated time series over the raw signal.  
This portion of data was chosen to highlight a potentially difficult section to decompose where the wave amplitude sharply decreases over time. 
We expect the decomposition to perform best when the wave energy is large relative to that of the turbulence. 
While we see strong agreement with the wave signal during the first half of the plot, we see the mismatch increases once the wave amplitude decreases in the second half of the plot, however the decomposed wave velocity is still able to capture the transition moderately well and maintain the shape of the waves.  
When we later compare this decomposition to other methods, we find it has similar performance to the EEMD and SWT methods because this is the dataset with the largest relative wave energy, i.e., it is theoretically the easiest signal to decompose.  
Additionally, Table \ref{tab:summaryDMDparams} indicates that this is the dataset with the highest relative wave intensity $\tilde{u}_{\text{rms}}/u'_{\text{rms}}$ and also the lowest $r/n$ values. 
This indicates that the wave motion present in the signal can be captured using a smaller number of modes relative to the total rank.

\subsection{Laboratory data decomposition}

\begin{figure}[htpb]
    \centering
    \includegraphics[scale=1]{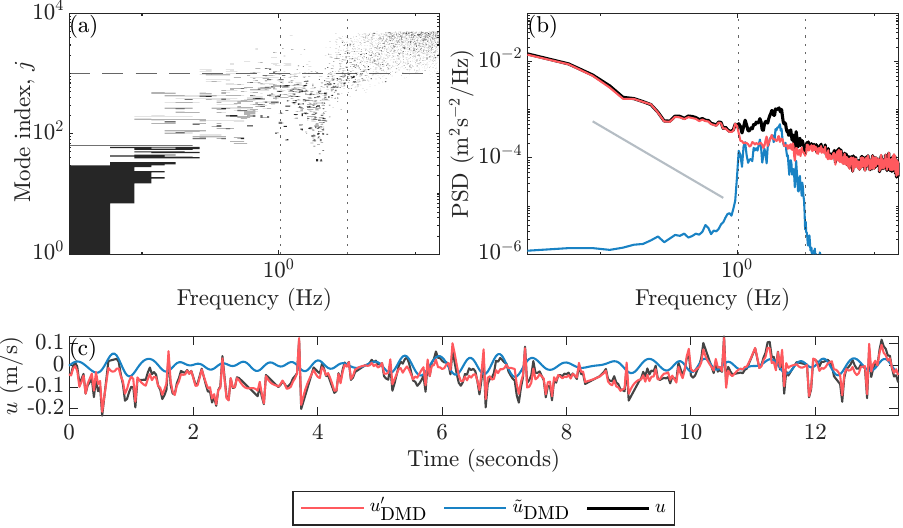}
    \caption{DMD wave-turbulence decomposition of $u$-component velocity of laboratory data. (a) POD spectrum of time-delayed matrix $\bm{\mathsf{H}}$. (b) Power spectra of the original data and DMD-separated wave and turbulence. \rev{A gray -5/3 slope line is plotted for reference.} (c) A portion of the time-series of the original and decomposed signals. The wave motion was reconstructed using 506 DMD modes.}
  \label{fig:LabDataDMDdecomp_u}
\end{figure}

\begin{figure}[htpb]
    \centering
    \includegraphics[scale=1]{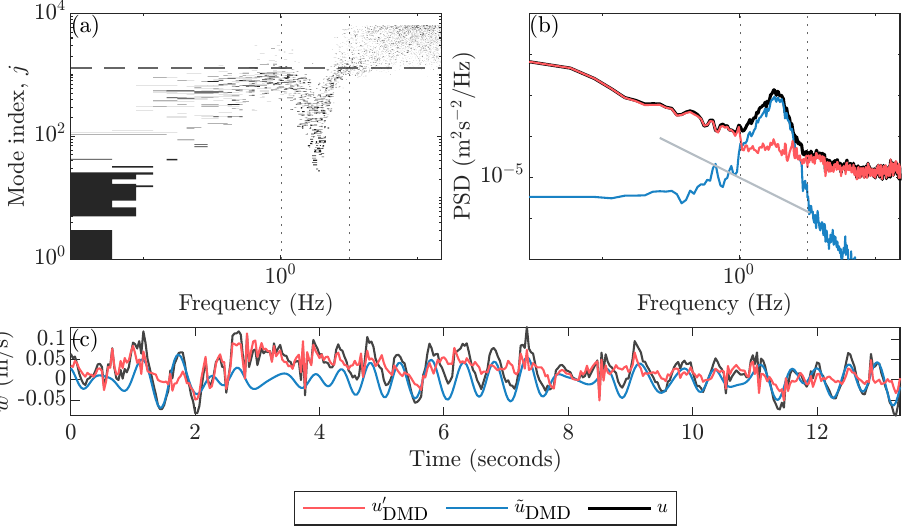}
    \caption{DMD wave-turbulence decomposition of $w$-component velocity of laboratory data. (a) POD spectrum of time-delayed matrix $\bm{\mathsf{H}}$. (b) Power spectra of the original data and DMD-separated wave and turbulence. \rev{A gray -5/3 slope line is plotted for reference.} (c) A portion of the time-series of the original and decomposed signals. The wave motion was reconstructed using 732 DMD modes.}
  \label{fig:LabDataDMDdecomp_w}
\end{figure}

The laboratory data represents a case with weak wave energy relative to the turbulence, and therefore we expect it to be the most challenging to decompose. 
The data is also more noisy than the field and synthetic data, but it is also the longest dataset of the three. 
We start by considering the POD spectra of both $u$ and $w$ (Figs. \ref{fig:LabDataDMDdecomp_u}a and \ref{fig:LabDataDMDdecomp_w}a) which show that the wave energy is contained within the frequency range of $1$ to $3$ Hz. 
What is interesting in these POD spectra, unlike in those of the synthetic and field data, is that the low number modes only contain low frequency turbulence, and that the wave energy is contained in intermediate modes.  
Based off the POD spectra, we truncate at $r=1000$ and $r=1300$ modes for $u$ and $w$ respectively. 

The fully decomposed spectra are plotted in Figs. \ref{fig:LabDataDMDdecomp_u}b and \ref{fig:LabDataDMDdecomp_w}b. We see that the DMD method separated and removed most of the wave energy from the turbulence in both cases.
First considering the  $u$ spectrum (Fig. \ref{fig:LabDataDMDdecomp_u}b), we see that the majority of the wave peak is removed from the turbulence spectrum, and that the wave spectrum has less energy than the turbulence spectrum at almost all frequencies, confirming that this is a turbulence-dominated flow. 
The time-series reconstruction in Fig. \ref{fig:LabDataDMDdecomp_u}c again shows the small wave velocity amplitude compared to the turbulence fluctuations. We also see some noise spikes in the data which have been isolated to the turbulence time series. 

Next, we consider the $w$ decomposition in Fig. \ref{fig:LabDataDMDdecomp_w}b where we observe a cleanly separated turbulence and wave spectra. Note the difference between this and the $u$ spectra, specifically how they have similar wave energy but the $w$ spectra has much weaker turbulence.   
The time-series reconstruction in Fig. \ref{fig:LabDataDMDdecomp_w}c show that the separated waves closely follow the trend of the raw data, and that both the high and low frequency turbulence fluctuations are isolated to the turbulence time series.
In Table \ref{tab:summaryDMDparams}, we see that this is the dataset with the lowest wave intensity relative to the turbulence.  This low wave intensity is a challenge for our decomposition, however by using a long enough time-series and a high $r/n$ truncation, the DMD method is able to give a better separation when compared with other methods. 

In order to further assess the decomposition results, \rev{we compare the mean Reynolds stress $\langle u'w'\rangle$ from the raw signal with those from the decomposed turbulence and wave components. 
We expect the Reynolds stress associated with waves to be significantly lower than that generated by turbulence. 
Consistent with this expectation, our analysis shows that the wave Reynolds stress is over an order of magnitude smaller than the residual turbulence Reynolds stress. 
We find the Reynolds stress values to be $-5.1 \times 10^{-4}$, $-4.5 \times 10^{-4}$, and $-0.19 \times 10^{-4}$ (m/s)$^2$ for the raw, decomposed turbulence, and decomposed wave signals respectively. This confirms that the wave momentum flux is inefficient compared to that of the turbulence, and that the DMD decomposition can successfully isolate wavelike motions from turbulence.}

\begin{figure}[htpb]
    \centering
    \includegraphics[scale=1]{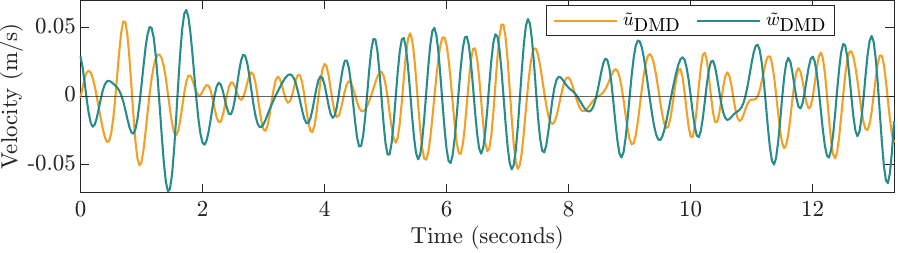}
    \caption{The $u$ and $w$ velocity components of DMD-separated wave time series of the laboratory data.} 
    \label{fig:LabData_phaseShift}
\end{figure}

Additionally, we visualize the time series of $\tilde{u}_{\textrm{DMD}}$ and $\tilde{w}_{\textrm{DMD}}$ together in Fig. \ref{fig:LabData_phaseShift}. 
We expect the horizontal and vertical wave velocities to have a 90° phase shift in time and for them to have similar amplitudes given that they were deep-water waves (short waves relative to the depth of the tank).
The initial section of Fig. \ref{fig:LabData_phaseShift} shows a disagreement between both the expected wave magnitudes and the phase shift, which might be caused by the presence of strong turbulence; however, the rest of the time series shows good agreement with the expected wave characteristics from theory. Throughout the time series, the $\tilde{u}_{\textrm{DMD}}$ velocity amplitude is slightly underpredicted when compared with the $\tilde{w}_{\textrm{DMD}}$ signal; this is likely due to the $u$ decomposition underperforming due to the stronger turbulence.
Overall, we find that the DMD method is able to decompose the signals fairly well, even given a noisy dataset with relatively weak, irregular waves under intermittent wave-breaking.

\subsection{Comparison with other decomposition methods}\label{sec:compare}

\begin{figure}[htpb]
    \centering
    \includegraphics[scale=1]{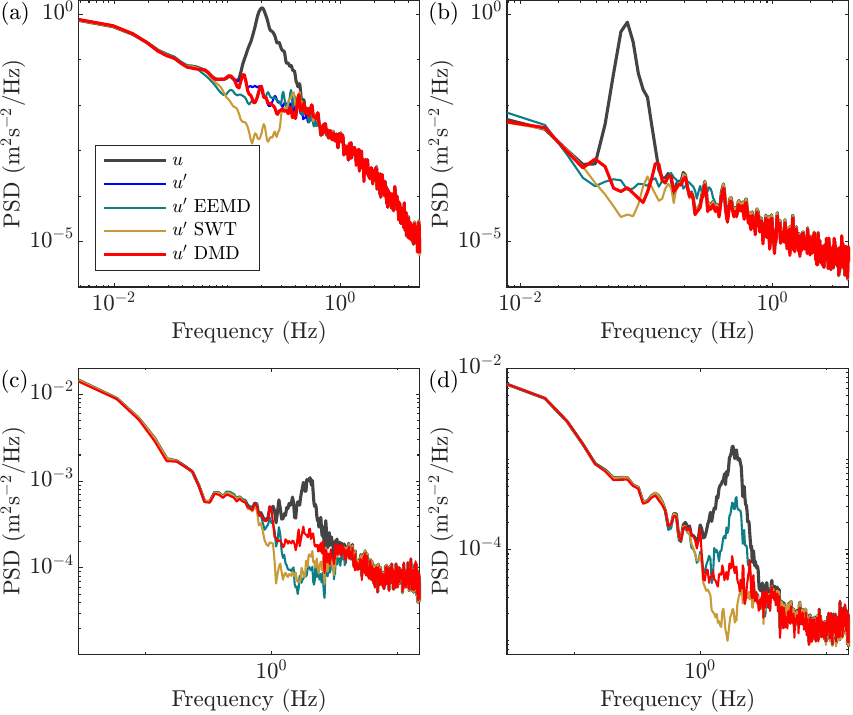}
    \caption{Comparison of modal decomposition methods, showing the spectra of the original data and the decomposed turbulence data. (a) Synthetic data $u$ velocity, (b) field data $u$ velocity, (c) and (d) laboratory data $u$ and $w$ velocity, respectively.} \label{fig:DataSummarySingle}
\end{figure}

We compare our decomposition method to the EEMD and SWT, two other modal decomposition techniques that work with a single time series. To compare the methods, we plot the separated turbulence spectra using each method for each dataset in Fig. \ref{fig:DataSummarySingle}. 
Across all four plots, we see that the SWT tends to excessively remove energy from the turbulence in the wave range, causing significant dips in the turbulence spectra. In contrast, the EEMD generally performs as well as the DMD method for the synthetic and field data in Figs. \ref{fig:DataSummarySingle}a and \ref{fig:DataSummarySingle}b, respectively. \rev{We note that the EEMD has caused some turbulence energy redistribution, e.g., adding energy into the turbulence in the field data as seen in Fig. \ref{fig:DataSummarySingle}b between 0.2 to 0.4 Hz.} Finally, the laboratory data (Figs. \ref{fig:DataSummarySingle}c and \ref{fig:DataSummarySingle}d) is the hardest to decompose due to its weak, irregular waves, and that is where we clearly see that the DMD outperforms the other two methods. The DMD filtered spectra is able to closely follow the expected slope of the turbulence spectra for both the $u$ and $w$ velocity spectra. 

The methods show the most similar decomposition for the field data, which is the dataset with the highest wave intensity. This is unsurprising and suggests that the stronger the wave intensity, the easier the signal should be to decompose using any method. Even though the synthetic data has exactly no wave-turbulence interactions which should in theory mean it is relatively easy to decompose, its lower wave intensity causes issues for the SWT method. Finally, the laboratory data has the lowest wave intensity, and this is where the benefits of the DMD method are the clearest. One reason the SWT may underperform is because it determines its basis \emph{a priori}; the method tries to fit the data to a wavelet basis, which might not always be a good descriptor of the dynamics of the system. Alternatively, the EEMD uses IMFs as an \emph{a posteriori} basis, that is, they adapt to the data being analyzed. However, it seems that the IMFs might not be separating the waves from the turbulence completely, resulting in mixed modes which contain both turbulence and wave energy. The utility of this method is further hindered by the fact that the number of IMFs is determined by the EEMD algorithm which removes a tuning parameter.

\begin{figure}[htpb]
    \centering
    \includegraphics[scale=1]{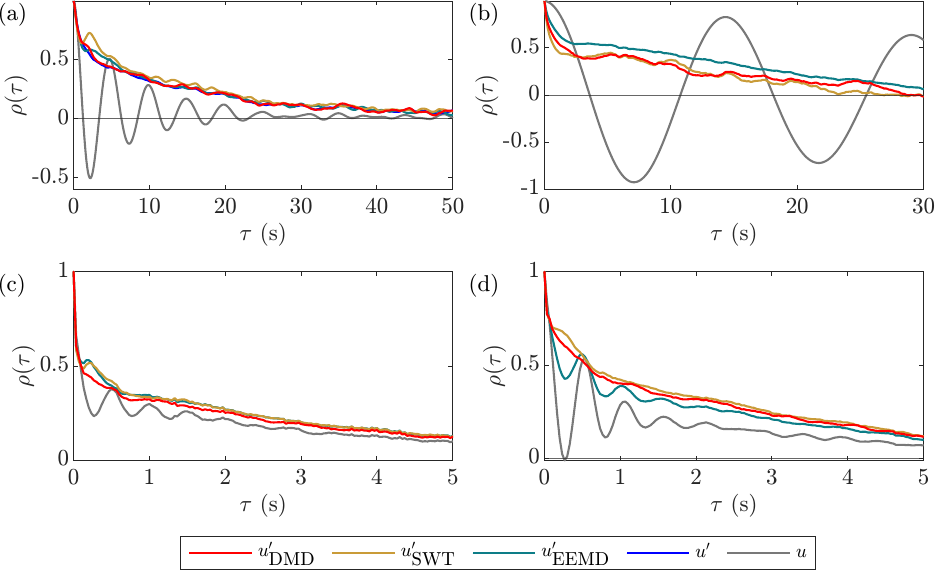}
    \caption{Normalized autocorrelation of raw ($u$), DMD-filtered turbulence ($u_{\text{DMD}}$), EEMD- and SWT-filtered turbulence ($u_{\text{EEMD}}$ and $u_{\text{SWT}}$, respectively) signals. (a) Synthetic data $u$ velocity, the black line is the true turbulence, $u'$. (b) Field data $u$ velocity. (c) and (d) Laboratory data $u$ and $w$ velocity, respectively.} \label{fig:DataAutocorr}
\end{figure}

We further compare the methods by computing the normalized autocorrelations $R(\tau)$ of the filtered turbulence velocity across the different datasets, as shown in Fig. \ref{fig:DataAutocorr}. 
The raw autocorrelations, plotted in gray for reference, show a clear periodic wave signal across all datasets. 
Successfully isolating the turbulence from the waves would remove all time periodicity and preserve the exponential decay expected as the turbulence fluctuations decorrelate over time. 
We see that in most cases, the methods are all able to remove the majority of the wave signal. 
With respect to the synthetic data (Fig. \ref{fig:DataAutocorr}a), we know what the true turbulence signal is, and we see that while all the methods follow the true turbulence autocorrelation (shown with a blue line) closely, the SWT performs the worst, as shown by the small peak in the signal after the initial decrease. 
In Fig. \ref{fig:DataAutocorr}b, the field data autocorrelations all remove the wave, and it is unclear which performs best here from visual inspection. Recall that the EEMD method had added energy into the turbulence spectra, so it's likely that its higher autocorrelation values are inaccurate. 
Finally, in the laboratory data (Figs. \ref{fig:DataAutocorr}c and \ref{fig:DataAutocorr}d), we clearly see the DMD method performs the best. It is able to remove all wave peaks while still capturing the long-time temporal decorrelation of the turbulence. 
This analysis clearly shows the strength of the DMD-method across a variety of datasets and demonstrates how it can be used for more than just spectral analysis, particularly under conditions that pose challenges to other decomposition methods.

\section{Conclusion}
\label{sec:Conclusion}
While many wave-turbulence decomposition methods have been developed over the years, they are typically limited in their use case. For example, they may be restricted to the spectral domain precluding a time-series construction, they may require multiple synchronized measurements, or they may have restrictive assumptions such as no wave-turbulence interactions.   
In many scenarios, this assumption can easily break, especially because waves and turbulence do interact across different time scales \citep{Guo_2013}. 
Given these challenges, newer modal decomposition methods have been developed, such as the EEMD and the SWT. 
These methods still have their drawbacks.  For example, the SWT uses a predetermined basis to project the data that is not flexible for highly nonlinear data whereas the EEMD introduces noise as a filter to differentiate scales in the time series \rev{which can lead to a more accurate separation of the time series' intrinsic modes. 
However, this process can still result in shifts in how the energy is distributed across the IMFs due to the artificial noise introduced to the decomposition process.}

These approaches are not optimal for highly nonlinear data, resulting in either poor or excessive wave energy separation. 
To address these limitations, we developed a new wave-turbulence decomposition method using DMD, leveraging its data-driven adaptive basis and minimal assumptions. 
Our method applies DMD to a single time-series,  assisted by a time-delay embedding. 
It is able to isolate the wave components without significantly affecting the turbulent signal at similar frequencies. 
In the case of velocity data, we show how the method is able to recover the inertial range scaling below the wave peak. 
The main assumptions that underlie this method are that the waves and turbulence can be separated and that waves are the most coherent feature of the time-series, which are reasonable assumptions considering the dynamics of the system.
We applied our method to synthetic, field, and laboratory data which covered a range of wave intensities. 
Our method clearly outperformed the SWT and EEMD when applied to the synthetic and laboratory data; whereas all methods performed similarly when applied to the field data. 
This may be due to the fact that the field data had the largest wave intensity, and therefore the wave motion was relatively easy to isolate from the turbulence across all of the methods. 
\rev{A sensitivity analysis on the effect of wave intensity shows that the DMD method performs best when the wave energy is equal or greater than that of the turbulence. And given that the decomposition relies on isolating the coherent wave motion, the decomposition performs the worst for low wave intensity because the method is unable to identify the wave motion once it is overshadowed by stronger turbulence fluctuations.}  

The proposed method does require some manual tuning: specifically the user needs to input the shape of the time-delay matrix, the rank truncation, and the wave frequency range of interest. 
However, we argue that the parameters are physically related to the signal, especially compared to other methods such as the SWT and EEMD which have somewhat arbitrary tuning parameters.
The rank truncation is discernible in the POD spectrum of the time-delayed input matrix, and the wave-frequency range is clearly visible in the power spectrum of the signal. 
While the number of time delays is less clear, our sensitivity analysis shows that it is not as important to the accuracy of the decomposition, especially when the time-series length is long enough. 
We recommend future work to examine this parameter more closely and relate it directly to the wave spectrum.  
The reported parameters in Table \ref{tab:summaryDMDparams} can be a starting point for users of the method.
Overall, the proposed DMD-based decomposition method is successful at decomposing wave and turbulence motion with minimal tuning and only one component of data, which is valuable for time-series measurements of flow parameters. 

We have demonstrated that DMD is powerful for this application, and some fine tuning may need to occur as it is adopted for this problem.
Future work can focus on more robust DMD schemes and alternative ways of determining the number of modes necessary to reconstruct the wave motion.
A few examples include the extended DMD \citep{Williams_2015} which attempts to circumvent certain constraints of the (linear) DMD technique when decomposing data from nonlinear systems, and sparsity-promoting DMD \citep{Jovanovic_2014} that can reduce the complexity of the calculations by identifying a smaller set of important DMD modes. 


\clearpage
\acknowledgments
We would like to thank Steve Brunton and Caroline Cardinale for helpful comments and insights. 
We acknowledge support from the U.S. National Science Foundation under grant No. CBET-2237550 and the Link Foundation Ocean Engineering and Instrumentation Ph.D. Fellowship.
All authors declare that they have no conflicts of interest.

%
%
\datastatement
The field data analyzed in this article are available in \citet{Reimers_2021} and its supplementary materials.
The synthetic and laboratory data that support the findings of this article, as well as the software used for the analysis, are openly available at \url{https://github.com/DiBenedettoLab/Wave-Turbulence_DMD}.

\bibliographystyle{ametsocV6}
\bibliography{references}

\end{document}